\documentclass[fleqn,10pt]{wlscirep}
\usepackage[utf8]{inputenc}
\usepackage[T1]{fontenc}
\usepackage{newunicodechar}
\usepackage{tablefootnote}
\newunicodechar{ʼ}{'}

\newcommand{\kms}{km~s$^{-1}$}

\newcommand{\aap}{{Astron. Astrophys.}}
\newcommand{\aaps}{{ Astron. Astrophys. Suppl.}}

\newcommand{\apj}{{Astrophys. J.}}
\newcommand{\apjl}{{Astrophys. J. Lett.}}
\newcommand{\apjs}{{Astrophys. J.  Suppl.}}

\newcommand{\grl}{{Geophys. Res. Lett.}}

\newcommand{\jgr}{{J. Geophys. Res.}}

\newcommand{\pasp}{{ Publ. Astron. Soc. Pac.}}

\newcommand{\solphys}{{ Sol. Phys.}}
 
\newcommand{\ssr}{{ Space Sci. Rev.}} 
\newcommand{\araa}{{ Annual Rev. Astron. Astrop.}} 
\title{Kinematic Relationship Between  Solar Extreme Ultraviolet Waves and Type II Metric Radio Bursts}

\author[1,*]{Ramesh Chandra}
\author[1]{Apoorv Dashora}
\author[2,3,*]{P. F. Chen}
\author[4,5]{Pooja Devi}
\affil[1]{Department of Physics, DSB Campus, Kumaun University, Nainital, 263001, India}
\affil[2]{School of Astronomy \& Space Science, Nanjing University, Nanjing 210023, Peopleʼs Republic of China}
\affil[3]{Key Laboratary of Modern Astronomy and Astrophysics (Nanjing University), Ministry of Education, Nanjing 210023, Peopleʼs Republic of China}
\affil[4]{Rosseland Centre for Solar Physics, University of Oslo, P.O. Box 1029, Blindern, N-0315 Oslo, Norway}
\affil[5]{Institute of Theoretical Astrophysics, University of Oslo, P.O. Box 1029, Blindern, N-0315 Oslo, Norway}

\affil[*]{rchandra.ntl@gmail.com; chenpf@nju.edu.cn}

\begin{abstract}
Solar extreme-ultraviolet (EUV) waves are large-scale disturbances that manifest as bright wavefronts, often during coronal mass ejections (CMEs). According to the magnetic field line stretching model, this phenomenon comprises two components: a fast-mode CME piston-driven shock wave and a slower, nonwave component. They are often associated with solar type II radio bursts. It is expected the radio source and the fast-mode EUV wave should come from different parts of the same shock, i.e., the CME piston-driven shock, and their speeds should be strongly correlated. To investigate this relationship, we utilized high spatiotemporal resolution observations from the Solar Dynamics Observatory in conjunction with radio data from the Radio Solar Telescope Network. 
Our analysis reveals that there exists a linear correlation between the EUV fast-mode speeds ($v_{euv}$) and the shock speeds derived from metric (m) type II radio bursts ($v_{radio}$), which is $v_{radio}=0.89v_{euv}+51$ \kms, with a correlation coefficient of 0.77. 
This strong correlation suggests that original coronal EIT waves, which are about three times slower than type II radio bursts, are not fast-mode waves, and it is misleading to map type II radio bursts to EIT waves in the literature.
\end{abstract}

\begin{document}
\flushbottom
\maketitle

\section*{Introduction}

The first observations of Moreton waves--large-scale disturbances propagating through the solar atmosphere at speeds of $\sim$~1000 \kms\ were reported in 1960s \cite{Moreton60}. These were identified mainly using H$\alpha$ blue and red wings, and occasionally using H$\alpha$ line-center observations. Three and a half decades after the discovery of Moreton waves, a large-scale coronal wave phenomenon was discovered in extreme-ultraviolet (EUV) wavelengths via the Extreme ultraviolet Imaging Telescope (EIT) aboard Solar and Heliospheric Observatory (SOHO) satellite. This phenomenon was named ``EIT waves'' after the telescope \cite{Thompson98}. The typical speeds of coronal ``EIT waves'' are within the range of 200--400 \kms, but sometimes they are over 700 \kms\ \cite{Thompson09}. After the Solar TErrestrial RElations Observatory (STEREO) and Solar Dynamics Observatory (SDO) were launched, the phenomenon was called EUV waves or large-scale coronal propagating fronts \cite{Nitta13}. With the higher cadence of the SDO data, it was revealed that the mean speed of the coronal EUV waves becomes 644 \kms, which is much larger than the values obtained via SOHO/EIT. In addition to the EUV wavelength, coronal propagating fronts are also visible in other wavelengths, including ultraviolet (UV), X-rays, and radio \cite{Warmuth05, white05,Chandra24}
The majority of EUV waves are associated with coronal mass ejections (CMEs) \cite{Chen06}. Moreover, the cases are reported where the EUV waves are accompanied by  solar flares\cite{Shen2018,Morosan2023}.

Although coronal EUV waves were initially believed to be coronal counterparts of chromospheric Moreton waves, i.e., they were thought to be fast-mode shock waves, observations revealed some features that cannot be explained by the fast-mode wave model. For example, some observations indicate the presence of stationary wave fronts situated at magnetic separatrices \cite{Delannee99, Del2000, Chandra16, Devi22}. A survey of the coronal EUV wave kinematics implied that there should be different classes of coronal EUV waves \cite{War11}. It was also noticed that the speed of a coronal EUV wave can sometimes be as small as $\sim$20 \kms, which falls far below the coronal sound speed, i.e., 150 \kms\ \cite{Zhukov09, Guo15}. Such a result cannot be interpreted by the fast-mode wave model. Furthermore, it was found, in the STEREO era, that the speeds of coronal EUV waves are anticorrelated with the local magnetic field, which is contrary to the prediction of the fast-mode wave model. All these observations challenge the proposal that coronal EUV waves are just fast-mode waves.

Several models have been proposed to explain coronal EUV waves. After they were discovered, EIT waves were soon widely believed to be fast-mode waves or shock waves  \cite{Thompson99, Wang2000, Warmuth01, Wu01, Vrsnak02}. However, the discovery of the stationary wave fronts rendered the fast-mode wave model inappropriate \cite{Delannee99, Del2000}. As a result, several other models were proposed \cite{Davey09,Warmuth10, Liu14, Chen16a, Chandra24}. 
To explain why coronal EIT waves are roughly three times slower than chromospheric Moreton waves, magnetic field line stretching model was proposed, i.e., coronal EIT waves are apparently propagating bright fronts which are produced when the overlying magnetic field lines are pushed by the erupting filament (or flux rope) to stretch up successively \cite{Chen02}. According to this model, the speed of a coronal EIT wave is about three times slower than the fast-mode wave in one event if the magnetic field lines overlying the filament are concentric semicircles. If the overlying magnetic field lines are vertically elongated, the corresponding EIT waves would be >3 times slower than the fast-mode wave. Later, it was confirmed that this model can naturally explain why EIT waves stop at magnetic separatrices to form stationary EIT wave fronts \cite{Chen05}. More importantly, this model predicted in 2002 that two coronal EUV waves should be observed in a single event if only the observational cadence is high enough, i.e., a faster one and a slower one, with the faster one being a fast-mode wave or shock wave and the slower one being coronal EIT wave. In this model, the fast-mode shock wave, not the EIT wave, corresponds to the coronal counterpart of the associated chromospheric Moreton waves. Later, both components of coronal EUV waves were confirmed in observations \cite{Chen11, Asai12, Chandra16, Chandra22, Devi22} and references cited therein.

There are other models to explain the mismatch of coronal EIT waves and chromospheric Moreton waves, such as successive reconnection model \cite{Attril07, Cohen09} or slow-mode wave model \cite{Davey09}. However, it is not clear how these models can explain the three-fold relationship between coronal fast-mode waves and EIT waves.

According to Uchida's model for Moreton waves \cite{Uchida68}, solar flares generate a shock wave propagating in the corona, with the top part generating type II radio bursts and the skirt of the shock front perturbing the chromosphere to generate Moreton waves. After coronal EIT waves were discovered, they were initially thought to be the expected coronal shock waves in Uchida's model. If so, the speeds of the EIT waves and type II radio bursts should be strongly correlated and similar. Several studies tried to address this relationship \cite{Klassen2000, Warmuth10, Nitta13, Long17}. However, these authors obtained complex and sometimes contradictory results.
For example, Klassen et al. \cite{Klassen2000} found that the majority of EIT waves were associated with metric type II radio bursts but there is no corelation between their speeds.
On the other hand, some later statistical studies reported significantly lower association rates i.e., 54\% \cite{Nitta13}, 22\% \cite{Muhr14}, and 40\% \cite{Long17}, between EUV waves and type II radio bursts. However, these studies revealed little correlation between EUV wave speeds and other properties of solar eruptions, e.g., the speeds of type II radio bursts, flare intensity, and CME speeds. 

These controversial results are possibly due to the fact that either only the slow component of coronal EUV waves (i.e., EIT waves) or a mixture of the fast and the slow components of EUV waves were considered. 
According to the magnetic field line stretching model \cite{Chen02}, which sometimes called as the hybrid model \cite{Nitta13}, it is the faster component of coronal EUV waves that shares the same origin as type II radio bursts. Therefore, in principle, the speed of the metric type II radio bursts should correlate with that of the fast component of the EUV wave. It implies that when we investigate the relationship between coronal EUV waves and type II radio bursts, it is crucial to distinguish the fast component EUV waves from the slow component waves, and consider the fast component EUV waves only, which is the purpose of this paper. 
 
 The paper is organized  as follows: Section ``Observations and Methods of Analysis'' presents the data sources and the analysis techniques adopted in our work. Section ``Results'' describes the spatial and the kinematical evolution of the studied EUV wave events together with their association with type II metric radio bursts and their connection with coronal mass ejections (CMEs) and flares. Finally, Section ``Discussions'' summarizes the main results of the work, which are briefly discussed.

\section*{Observations and Methods of Analysis}

For the present study, we analyzed the data from the following instruments:
\begin{enumerate}
    \item {For the EUV wave analysis, we took the data from the Atmospheric Imaging Assembly (AIA) \cite{Lemen12} on board the Solar Dynamics Observatory (SDO) \cite{Pesnell12} satellite and the Extreme Ultraviolet Imager (EUVI) on board the Solar TErrestrial Relations Observatory - Ahead (STEREO-A) satellite \cite{Howard08}. The AIA observes the full Sun in seven EUV, two UV wavelengths, and one white light channel. The pixel size and cadence of the AIA data is 0.6$''$ and 12 sec, respectively. The STEREO pixel size and cadence are is 1.6$''$ and  2.5 min, respectively. For the current study, we used the AIA data in 193 \AA\ and STEREO--A EUVI 195 \AA\ due to the clear visibility of the EUV wave in this channel.
    First, we examined all EUV waves provided at {\url {https://www.lmsal.com/nitta/movies/AIA_Waves/oindex.html}} for the current Solar Cycle 25 (from December 2019 to December 2025). This catalog contains movies of EUV waves in various EUV and UV wavebands of AIA. 
   To minimize projection effects on the calculated EUV wave speeds, we restricted our sample to events originating at longitudes greater than $\pm$ 60$^{\circ}$.
We selected the near limb events so that the EUV waves generally propagate toward the disk center, and some EUV waves propagate above the solar limb. In both cases the EUV wave speeds suffer less from the projection effects. In our data set eight events are backside events. Although these events originated on the backside, all of them were located near the solar limb.

\item{
For this study, we primarily selected those EUV wave events where both the fast-mode and nonwave components are clearly discernible. In cases where this distinction was unclear and only a single component was visible, we included the events only after confirming its identity as a fast-mode MHD wave.}

\item {Data for the associated type II metric radio bursts were obtained from the Radio Solar Telescope Network (RSTN) and E-Callisto network. The RSTN monitors the solar radio spectrum in a frequency range of 25--180 MHz using a global network of four observatories: Learmonth, San Vito, Palehua, and Sagamore Hill.
Here, we would like to mention that the onset of m-type II bursts are constrained by the upper observational frequency in the RSTN network because some of the m-type II bursts can start at higher frequencies than the RSTN observing range. }

\item  
    {To analyze the association between EUV waves and CMEs, we follow the Large Angle and Spectrometric Coronagraph (LASCO) CME catalog (\url {https://cdaw.gsfc.nasa.gov/CME_list})\cite{Gopalswamy2009}. In addition, the association between the EUV waves and GOES flare class is taken from  the NOAA solar active region (AR) summary reports (\url{https://www.ngdc.noaa.gov/stp/space-weather/swpc-products/daily_reports/solar_event_reports}).}
    }
   \end{enumerate} 

\begin{figure}[!t]
\centering
\includegraphics[width=\linewidth]{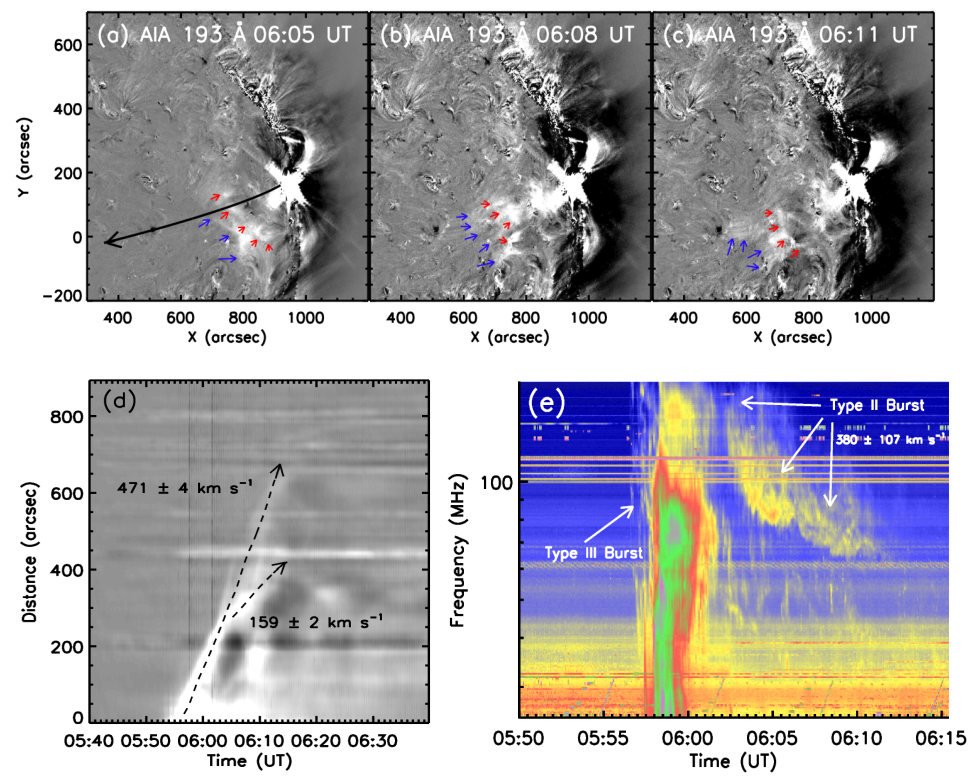}
\caption{An example of EUV wave on 2022 January 20, showing its evolution (a--c). The blue and red arrows indicate the fast component of EUV wave and the nonwave component, respectively. Panel `d' depicts the time-distance diagram along the artificial slice labeled with the black arrow in panel `a'. The fast-mode and nonwave components of the EUV waves are also shown in panel `d' with dotted lines. The associated type II metric radio bursts observed by RSTN network is plotted in panel~`e'.  }
\label{fig:two_component}
\end{figure}

\begin{figure}[t]
\centering
\includegraphics[width=\linewidth]{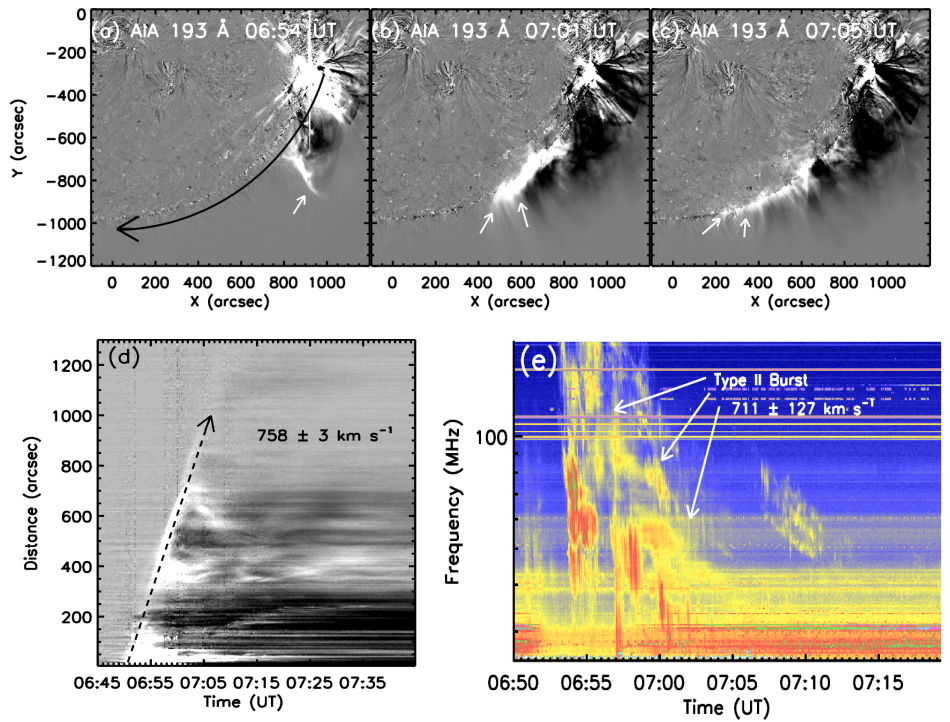}
\caption{Same as in Figure~\ref{fig:two_component}, but for the EUV wave event on 2024 February 16. This is an example where only the fast-mode EUV wave component is discernible.}
\label{fig:single}
\end{figure}

\begin{figure}[!ht]
\centering
\includegraphics[width=1.0\textwidth]{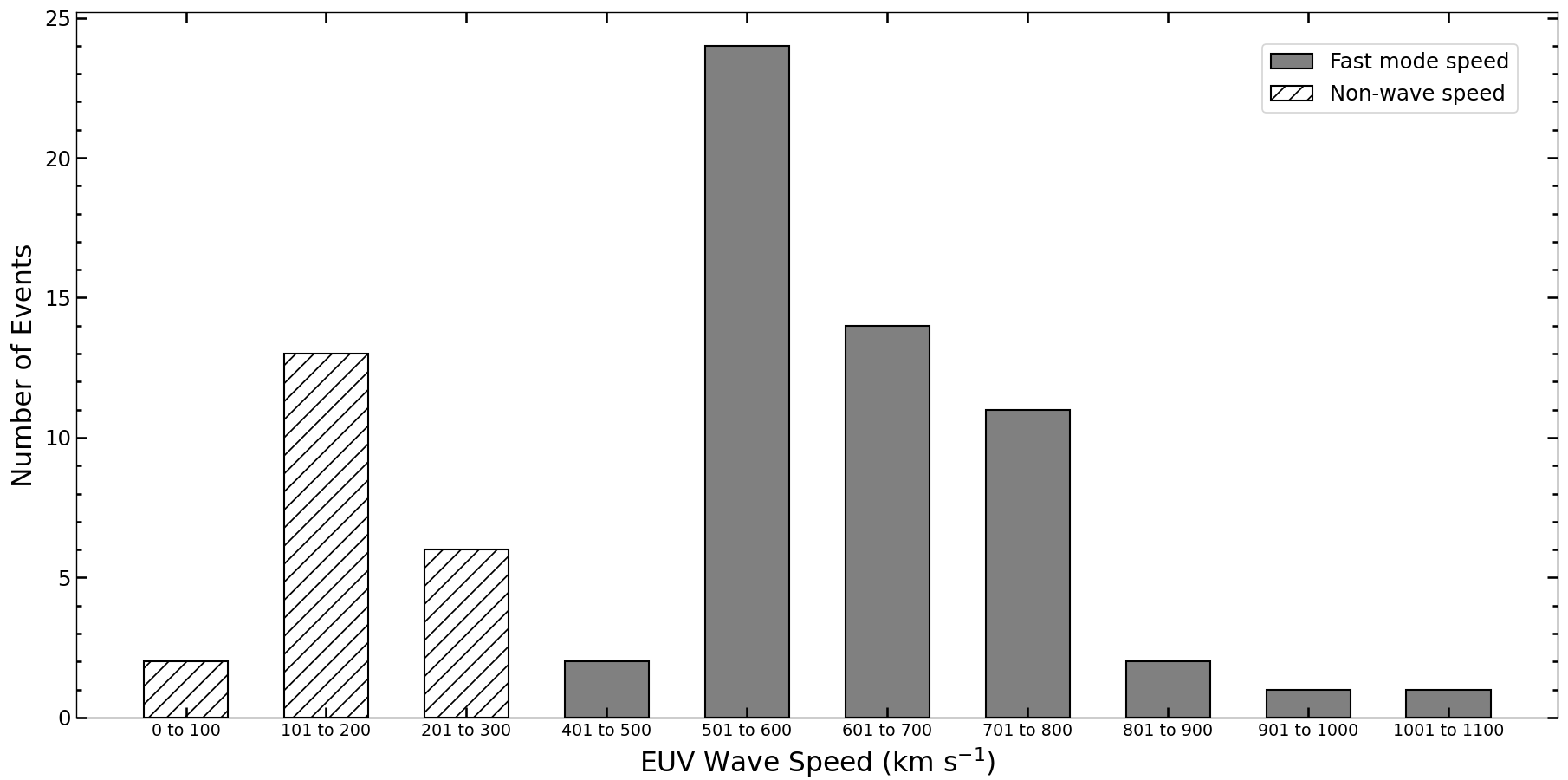}
\caption{Occurrence rates of the fast-mode EUV wave (gray bars) and the nonwave component EUV wave (hatched bars) with different speeds. The mean and median values of the fast-mode component waves are $\sim$628 and $\sim$610 \kms, respectively. For the nonwave component, the mean and median values are $\sim$179 and 172 \kms, respectively.}
\label{fig:EUV_speed}
\end{figure}

\begin{figure}[ht]
\centering
\includegraphics[width=\linewidth]{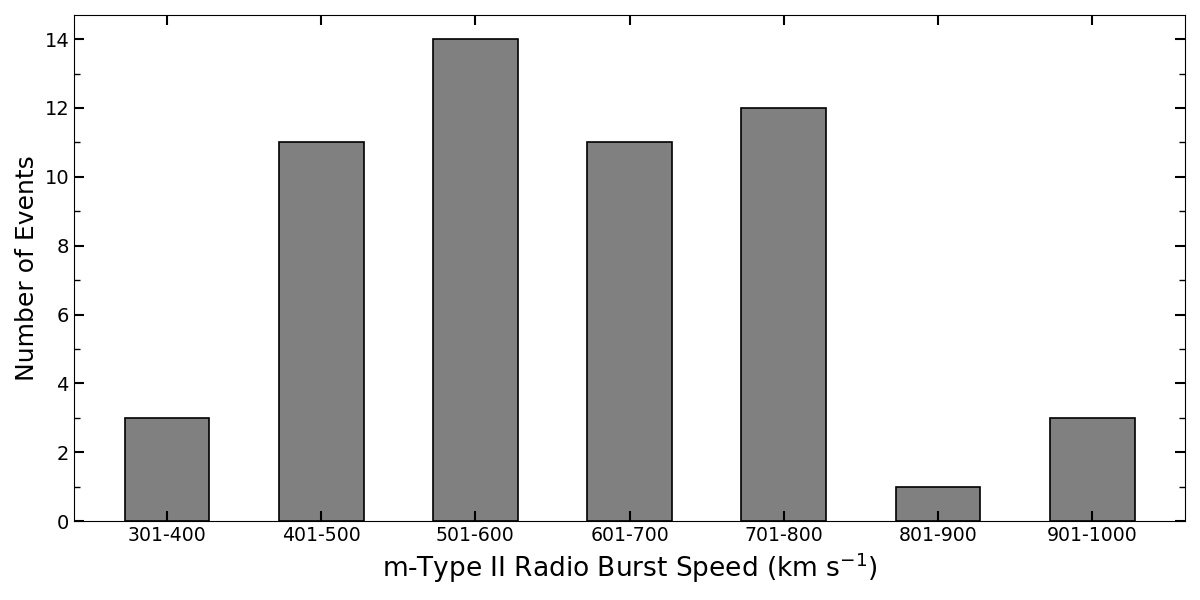}
\caption{Distribution of m-type II radio burst speeds. The mean and median values are $\sim$ 608 and $\sim$ 593 \kms, respectively. }
\label{fig:typeii_speed}
\end{figure}

 Data selection criteria and analysis procedures are described below: \\
 
  To distinguish the two components of the EUV waves, namely the fast-mode wave and the nonwave, we adopted the following criteria. 
We took artificial slices along all possible directions. After examining all these results, we selected the final slices in which both waves were most clearly discernible. When the distinction of the two components is unclear, the slice of highest velocity is selected. The starting points of the slices are the flare center. The width of the slice is 3.6$''$. 
 To distinguish the fast-mode and nonwave nature of the EUV wave, we adopted the benchmark explained by Chen \cite{Chen16a}. 
 According to Chen \cite{Chen16a}, when the speed of the EUV wave is greater than 500 \kms, it is treated as a fast-mode EUV wave. To create the time-distance diagram of the EUV waves, we first enhanced the visibility of the propagating wave front by creating base-difference images, i.e., subtracting a pre-event frame from subsequent images in the sequence.       
 Next, to compute the speeds of the EUV waves, we track the EUV wave fronts from the time-distance plot, for both fast-mode and slow component (wherever applicable), and fit a straight line in the data points. This straight line fit gives the average speed of the propagating front. 
       
The subsequent step involves searching for the associated m-type II radio bursts within the RSTN database. From this, we finalized a data set of 55 events, all of which exhibit clear m-type II radio emission features. 
To associate the type II and EUV wave events, we considered the type II lanes closest in time to the EUV wave for all events.
For the computation of the speeds of type II m-radio bursts, we applied four coronal density models namely: Newkirk (2  fold for AR)\cite{Newk61}, Saito \cite{Saito1970}, Leblanc \cite{Leblanc1998}  and hybrid density models \cite{Vrsnak2004}. First, we traced the type II radio burst and determined its fundamental frequency. For the harmonic emission, the fundamental frequency is half the emission frequency. Then, the corresponding plasma density of the radio burst source was calculated by the following formula:

\begin{equation}
  \label{eq:density}
  f{\rm [MHz]} = 9 \times 10^{-3} \times \sqrt{n} ~,
\end{equation}
\noindent
where $n$ is the corresponding plasma number density in units of cm$^{-3}$. 

Once the plasma density is determined, the height of the radio bursts, $R$, is computed using Equations (2--5). These equations represent the Newkirk, Saito, Lablanc, and hybrid density models, respectively.

\begin{equation}
\label{eq:Newkrik}
n [{\rm cm}^{-3}] = 4.2\times 10^4 \times N \times 10^{4.32/R}, ~{\rm where} ~ N=2 ~{\rm fold ~for~ AR}
\end{equation}

\begin{equation}
\label{eq:saito}
n [{\rm cm}^{-3}]=\frac {1.36\times 10^6} {R^{2.14}} + \frac{1.68\times 10^8}{R^{6.13}}
\end{equation}

\begin{equation}
n [{\rm cm}^{-3}] = \frac{3.3\times 10^5}{R^{2}} + \frac{4.1\times 10^6}{R^{4}} +  \frac{8\times 10^7}{R^{6}}
\label{eq:Lablanc}
\end{equation}

\begin{equation}
\label{eq:hybrid}
    n{\rm [10^8 ~cm^{-3}]} = \frac{15.45}{R^{16}} + \frac{3.16}{R^6} + \frac{1}{R^4} + \frac{0.0033}{R^2},
\end{equation}
\noindent
where $R$ is in units of solar radius.   

Different term of the hybrid density model shown in equation 5 are as follows:  The first term describes the AR corona, representing the five-fold Saito model \cite{Saito1970}. At low heights, this term roughly corresponds to the two-fold Newkirk model \cite{Newk61}, or five-fold Newkirk model \cite{Newkirk1967}. The fourth term represents the Leblanc model \cite{Leblanc1998}. The further details of this hybrid can be found in \cite{Vrsnak2004}.

Afterwards, the speed of the type II metric radio bursts, $v_{radio}$, is computed by 

\begin{equation}
\label{eq:speed}
  v_{radio} = \frac{dR}{dt} = \frac{R_{i+1} - R_{i}}{t_{i+1}-t_{i}},
\end{equation}
\noindent
where $t_i$ and $t_{i+1}$ are two moments, and $R_i$ and $R_{i+1}$ are the corresponding heights of the radio burst at the two moments.
Finally, we averaged the speeds derived using above mentioned four density models and the difference between them is regarded as the error bar of the speed of type II radio bursts.

The uncertainties of the computation of the speeds of type II radio bursts and the EUV wave linear propagation in the time-distance fits can be due to the visual selection of the points which can be 2 to 5 pixels. For each event, we selected several slices along different directions, and obtained different propagation speeds of the fast-mode EUV wave. The average of these speeds was taken to be the fast-mode EUV wave speed for the event, with the dispersion being the error bar.
To associate the EUV wave and the type II bursts, we linked them according to the nearest time of their occurrence. In case of two or more closeby m-type II bursts, we have measured speed of all possible type II bursts and finally averaged them.

\section*{Results}
We analyzed a sample of 55 EUV wave events from the current solar cycle, all of which occurred in association with type II metric radio bursts. The comprehensive kinematic parameters and the event details for this data set are listed in Table \ref{table}. 
The table provides a list of EUV waves with the day/time of their observation and the speeds of the fast-mode and nonwave components. When there is a `No' in the column of nonwave speed, it means that we did not find the nonwave component for this particular EUV wave event. Furthermore, this table provides the details of the associated type II radio bursts with their origination time and the computed speeds. The associated flare onset time and their GOES class are also given in the table. The last columns give the onset time of the associated CMEs and their speeds.

A primary work of our analysis is the identification of two distinct propagating EUV wave components, as exemplified in the event on 2022 January 20 shown in Figure \ref{fig:two_component}. Panels (a)--(c) illustrate the temporal evolution of both the fast-mode MHD wave and the slower nonwave component, marked by the blue and red arrows, respectively. The slice chosen for the time-distance analysis along the propagation direction of the EUV wave is shown in panel (a) with a black arrow, and the resulting time-distance diagram is shown in panel (d) of the figure. The time-distance plot clearly shows the temporal evolution of the fast-mode and nonwave components of the EUV wave event.  The fast-mode EUV wave is spatially and temporally correlated with the type II radio bursts recorded by the RSTN network (shown in panel e).

Our investigation reveals that these two (fast-mode and nonwave) components are clearly decoupled in 21 out of 55 events in our list. In the remaining 34 cases, however, the nonwave component was not observed. In these single-component events, the measured speeds range from 500 to 900 \kms, leading us to conclude that these waves really represent fast-mode MHD waves \cite{Chen16a}. An illustrative example of such a single-component event is the one on 2024 February 16, which is presented in Figure \ref{fig:single}, including its time-distance diagram and the associated type II radio burst. 

The time-distance diagram is further used to calculate the speeds of the EUV waves. We track the distance and time of the fast-mode wave and the nonwave component from the time-distance plots and fit any ridge with a straight line. The slope of the straight line gives the average speed of each wave.
In Figure~\ref{fig:EUV_speed}, we present a detailed analysis of the speed distributions for both the fast-mode EUV wave and the associated nonwave components. Our observations indicate that fast-mode wave fronts propagate with speeds ranging from 400 to 1100 \kms, yielding a calculated mean value of 628 \kms. In contrast, the nonwave components exhibit much slower kinematic profiles, with speeds varying from only a few \kms\ to a maximum of 300 \kms. The mean speed for the nonwave component is 179 \kms. 

Next, we compute the speed of the associated type II radio bursts using the Newkirk, Saito, Lablanc, and hybrid density models, and used the average of these computed speeds. Figure~\ref{fig:typeii_speed} illustrates the speed distribution of associated type II radio bursts. Their speeds vary from 350 \kms\ to $\sim$ 1000 \kms. The majority of type II events (14 events) lie in the speed range of 501 to 600 \kms\ with a mean speed of 608 \kms.

Based on the hybrid model of EUV waves \cite{Chen02}, the fast-mode EUV wave component originates from the CME piston-driven shock. Therefore, the speeds of fast-mode EUV waves and type II radio bursts should strongly correlate with each other.
In order to verify such a conjecture, we performed a correlation analysis between the propagation speeds of the fast-mode EUV waves and the speeds of the associated type II radio bursts. 

\begin{figure}[t]
\centering
\includegraphics[width=0.8\linewidth]{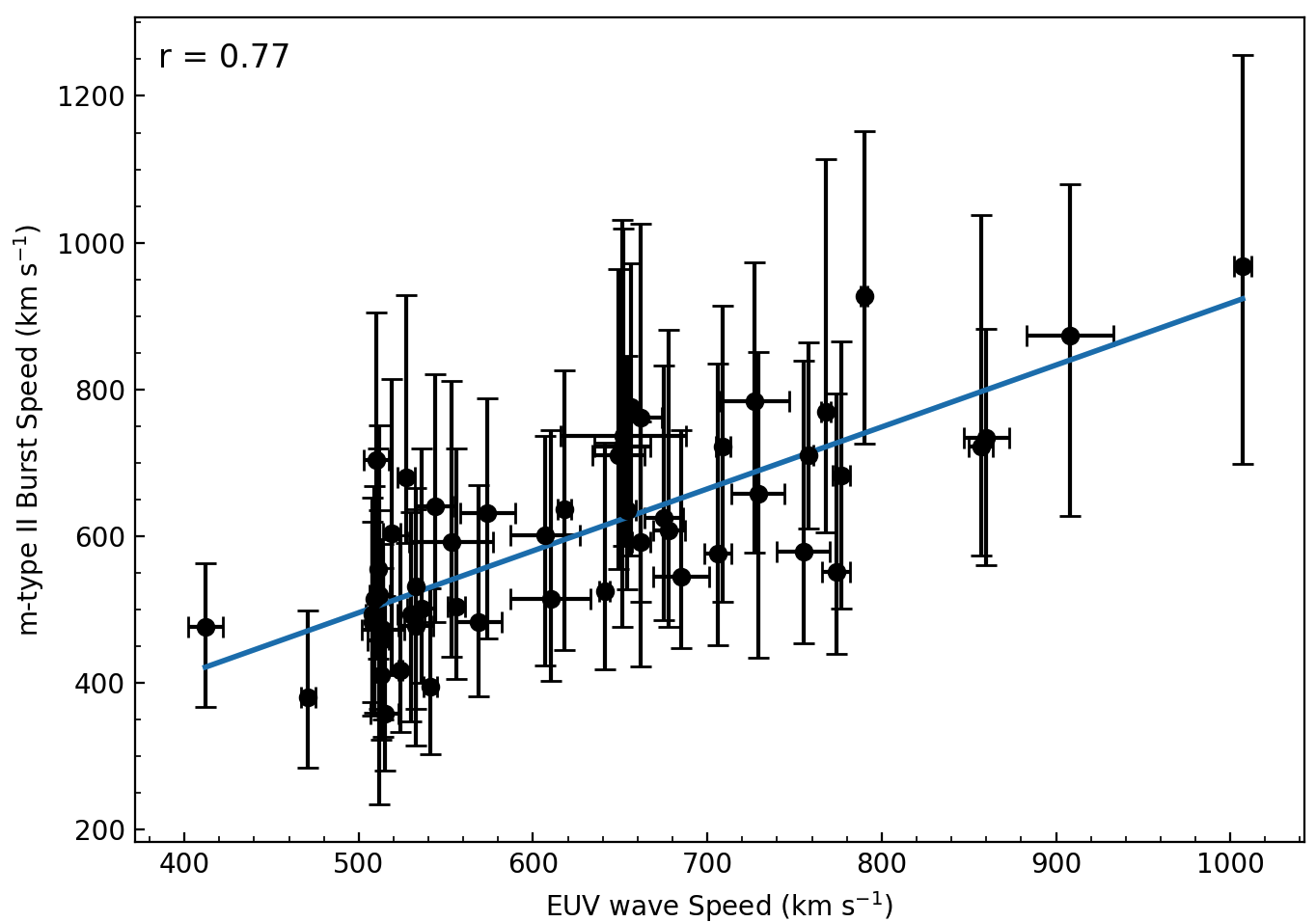}
\caption{Relationship between the fast-mode EUV wave speeds and m-type II radio burst speeds, which is overlaid with a linear fit {\bf (blue line)}. The error bars of the EUV wave speeds and type II radio burst speeds are also drawn. The value of the calculated correlation coefficient ($r$) is displayed at the top-left corner of the plot. }
\label{fig:Correlation}
\end{figure}

As illustrated in Figure \ref{fig:Correlation}, the two quantities show a significant correlation. Linear regression analysis was applied to quantify their relationship. Our results reveal a strong positive correlation, which yields a Pearson correlation coefficient of r=0.77. More importantly, we fit the data points with a linear function, which turns out to be $v_{radio}=0.89v_{euv}+51$ \kms.

\begin{figure}[ht]
\centering
\includegraphics[width=0.9\linewidth]{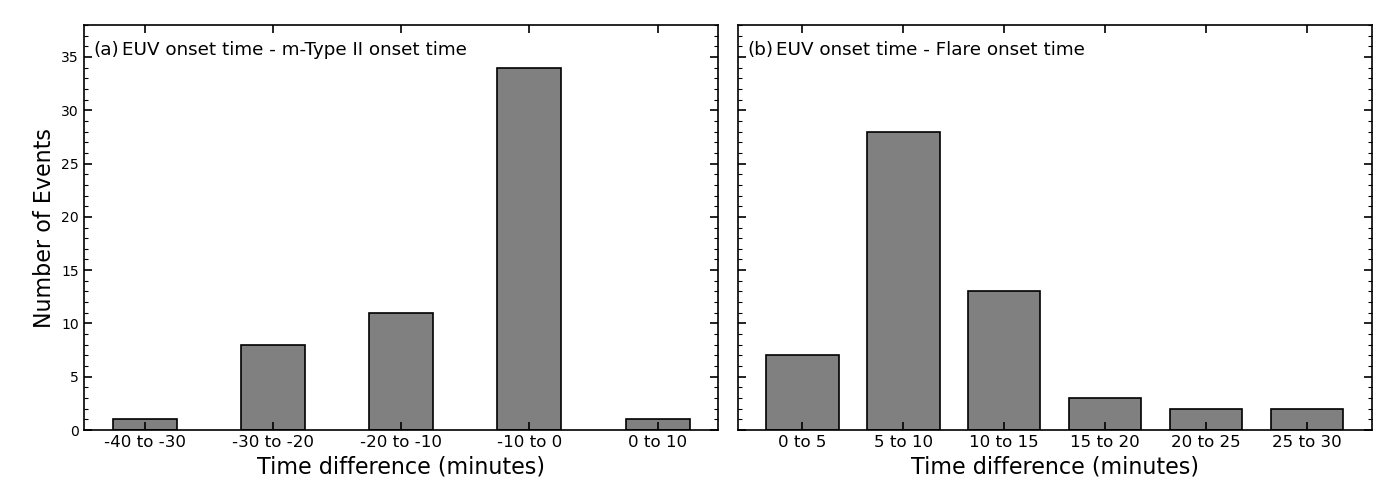}
\caption{Distributions of the time delay (relative to the fast-mode EUV waves) of type II radio bursts (panel a), and solar flares (panel b).}
\label{fig:time_difference}
\end{figure}

All EUV wave events in our list are associated with flares ranging from GOES C to X classes, as presented in Table \ref{table}.  We find that among the 55 wave events, 9 flares are of the C-class, 24 are of the M-class, and 14 are of the X-class. In addition, 8 events originate from the back side of the Sun. We note the onset times of the flare and compare them with those of EUV waves (panel (b) of Figure \ref{fig:time_difference}). Similarly to the association with flares, all events are found to be associated with CMEs. The speeds of associated CMEs range from 221 to 2782 \kms, with a mean speed of 1101 \kms. The majority (44) of CMEs are fast CMEs with speeds >500 \kms\ and only 11 CMEs are slow events with speeds <500 \kms. The speeds of the CMEs are taken directly from the LASCO CME catalog. We also took the angular width of the CMEs from the catalog (see Table \ref{table}). Among the sample, 31 (56.3\%) events are halo CMEs, 18 (32.7$\%$) are partial halos (whose angular width >120$^{\circ}$), and only 6 (11\%) are narrow CMEs ($<$ 120$^{\circ}$).

Figure \ref{fig:time_difference} illustrates the temporal relationships between the various eruptive phenomena by exhibiting the time offsets between the onsets of EUV waves, type II radio bursts, and solar flares. 
The temporal offset between the initiation of EUV waves and the onset of type II radio bursts (Panel a) reveals a tight synchronization between these two phenomena. As shown in our analysis, the onset time difference varies from -40 to +10 min, with a significant majority of events being within the -10 to 0 min offset range. This negative offset indicates that, in most cases, the EUV wave precedes the detection of the metric radio burst by several minutes.
The delay between the EUV wave onset time and GOES flare onset varies from 0 to 30 min, and the maximum occurs in the range from 5 to 10 min (panel b). All the temporal offsets in this panel are positive, which means that EUV waves always start after the flare onset.

\section*{Discussions}

The association between coronal EUV waves and type II radio bursts has been a topic of debate. In order to obtain a clear interpretation, in this study we selected 55 events located at greater than $\pm$60$^{\circ}$ longitude range that clearly display two components of EUV waves or contain at least the fast-mode EUV wave. This criterion ensures a reliable comparison between the fast-mode EUV wave speed and the type II radio burst speed. This is important because only the fast-component, not the slow component, of the EUV wave originates from the same shock that generates type II radio bursts. 
The prime objective of this study is to investigate the association between the fast--mode EUV wave speed and the associated type II radio burst speed. 
The main results of our study are summarized as follows:
\begin{itemize}

\item The speeds of the EUV fast-mode component strongly correlate with type II radio burst speeds, showing a correlation coefficient of 77\%.

\item In our selected data sets, $\sim$38\% (21/55) of the EUV wave events show two clear components. However, $\sim$62\% (34/55) events display only the fast-mode EUV wave component.

\item All events are associated with CMEs and solar flares. Most of the associated CMEs are fast and halo events, and most of the associated flares are M-class.

\end{itemize}

The association of EUV waves with type II radio bursts and the correlation between their speeds have been an unclear topic.
\cite{Klassen2000} demonstrated a high association rate, i.e., $\sim$90\%, between EIT waves and type II bursts, implying that they are intimately associated with CMEs rather than solar flares as even X-class flares without CMEs cannot excite EIT waves \cite{Chen06}. However, they revealed that the derived speeds of type II radio bursts are typically three times faster than those of EIT waves. On the other hand, numerous studies claimed that type II radio bursts are cospatial with EUV waves \cite{Khan2002, Zucca2014, Koukras2020, grec22, Vasanth2024, Zucca2025}.

We think that the contradictory results of the literature are due to that the past studies on this topic did not consider separating the slow and fast-mode components of EUV waves. In particular, before the SDO/AIA era, the low cadence of EUV imaging telescopes hindered observers from detecting the fast-mode EUV waves \cite{Chen11}. As a result, some authors chose the nonwave component EUV waves, others chose the fast-component EUV waves. Therefore, these studies were unable to reach a clear conclusion about the connection between EUV waves and their associated type II radio bursts, or about how their speeds are related. According to the standard flare/CME model \cite{Chen2011}, the fast-mode EUV waves are directly related to the type II radio bursts as both originate from different parts of the same piston-driven shock wave. In contrast, according to the magnetic field line stretching model \cite{Chen02}, the slow-component EUV waves are completely different from them in nature.

In this paper, we explicitly performed this separation, which represents a novel approach and enables a more reliable comparison between the two phenomena. We found that the fast-mode EUV waves are strongly correlated with type II radio bursts, with a correlation coefficient of 0.77. Their speeds are related with a linear function $v_{radio}=0.89v_{euv}+51$ \kms.
 The strong correlation between the fast-mode EUV wave speeds and the associated type II radio burst speeds can be interpreted by the hybrid model of coronal EUV waves \cite{Chen02}, which confirms the two components of EUV waves.
 It is the fast-component that is responsible for type II radio bursts, hence giving the strong correlation. 
 Similarly, \cite{Warmuth10} found a clear correlation between the speeds of chromospheric Moreton waves and type II radio bursts. The consistency between our results and theirs is simply due to the fact that the fast-component EUV wave is cospatial with the H$\alpha$ Moreton wave.

Our analysis revealed a 100$\%$ association rate, i.e., every EUV wave event is accompanied by both CME and solar flare. Moreover, it should be noted that some of the previous EUV wave events have been observed that do not have a CME association \cite{Shen2018, Morosan2023}. Among the associated CMEs, the majority (89\%) are halo or at least partial halo events. The flare magnitudes exhibit a broad distribution, ranging from GOES C-class events to major X-class eruptions. Specifically, 14 events are of X class and nine events are of C-class. In addition, eight events originate from the solar backside.

The ubiquitous presence of CMEs across our sample provides robust observational evidence for the coupled nature of these phenomena. This consistent association confirms the magnetic field line stretching model for coronal EUV waves \cite{Chen02, Chen09}, which posits that there are two types of coronal EUV waves, a faster one and the slower one. The faster one corresponds to the CME piston-driven shock wave, which generates both type II radio bursts and chromospheric Moreton waves, and the slower one corresponds to the coronal ``EIT waves'', which are cospatial with the CME frontal loop. In this paper, we did find that 34 of 55 events exhibit only the fast-component EUV waves, with the slow component nearly undetectable. The absence of the slow-component EUV waves was explained to be due to that the magnetic field lines are significantly forward-inclined \cite{Li2025}.

In summary, distinguishing between the fast-mode and nonwave components of EUV waves is crucial for space weather research. This separation is vital for establishing a clear physical link between EUV waves, solar energetic particles (SEPs), and subsequent geomagnetic activity along with the study performed here. It is the fast-mode wave component that is responsible for SEPs, not the slower nonwave component \cite{mite14}.

\section*{Acknowledgments}
{We thank the reviewers for their constructive comments and suggestions, which improved the quality of the paper.}
Authors thank the open data policy of NASA’s Solar Dynamics Observatory mission, Radio Solar Telescope Network, and e-Callisto data. We also acknowledge the LASCO-SOHO providing the archival data.

\section*{Author contributions statement}

{R.C. and P.F.C. conceived the research idea. R.C. and A.D. carried out the data selection and analysis. R.C. wrote the main part of the manuscript. P.D. and P.F.C. contributed to the revisions. All authors reviewed and approved the final manuscript}.

\section*{Funding}

 P.F.C. is financially supported by NSFC (12127901). R.C. and A.D. acknowledge support from the DST/SERB project number EEQ/2023/000214. P.D. acknowledges the support from the Research Council of Norway through its Centers of Excellence scheme, project number 262622.

\begin{table}[h]
\setlength{\tabcolsep}{3.5pt}
\centering
\footnotesize
\small
\caption{Various parameters of the EUV waves and their associated type II radio bursts and CMEs. The CME onset time means the time when a CME first appears in the LASCO C2 field of view. Type II speed is the average speed derived using Newkirk, Saito, Lablanc, and hybrid density models. Asterisk (*) represents the events ~observed by STEREO-A.}
\begin{tabular}{ccccccccccccc}
\toprule
Event & Date & EUV & EUV & \multicolumn{2}{c}{EUV Speed} & Type II & Type II & Flare & Flare & CME & CME  
& CME \\
No. & YYYY/MM/DD & Onset & Source & \multicolumn{2}{@{}c@{}}{($\mathrm{km\,s^{-1}}$)} & Onset & Speed & Onset & GOES & Onset & Speed & width  \\
      &      & Time & Location & Fast & Nonwave & Time & ($\mathrm{km\,s^{-1}}$)  & Time & Class & Time & ({$\mathrm{km\,s^{-1}}$}) & (in degree) \\
\hline
1  & 2020/11/29 & 12:50 & S23E95 & 556 & 193 & 12:58 & 505  & 12:34 & M4.4 & 13:25:49 & 2077 & Halo \\
2  & 2021/06/09 & 12:05 & N27W100 & 654 & No  & 12:01 & 635 & 11:57 & C1.7 & 12:24:05 & 441  & 65  \\
3  & 2021/06/23 & 06:47 & N16E90 & 574 & No  & 07:05 & 631  & 06:43 & C3.4 & 07:24:05 & 491  & 108 \\
4  & 2021/09/28 & 06:00 & S20W65 & 412 & 87  & 06:20 & 476  & 05:54 & C1.6 & 06:24:05 & 524  & 265 \\
5  & 2021/10/28 & 15:22 & S35W62$^{*}$ & 649 & 142 & 15:28 & 710  & 15:17 & X1.0 & 15:48:05 & 1519 & Halo \\
6  & 2021/11/01 & 01:07 & S27W60 &536 & No  & 01:29 & 502  & 00:57 & M1.5 & 02:00:06 & 753  & 274 \\
7  & 2022/01/12 & 04:20 & Backside &774 & No  & 04:25 & 551  & 04:10 & Backside & 04:36:05 & 1586 & Halo \\
8  & 2022/01/20 & 05:56 & N09W76 &471 & 159 & 05:57 & 380  & 05:41 & M5.5 & 06:12:06 & 1431 & 211 \\
9  & 2022/02/12 & 08:30 & S18W80 &508 & No  & 08:31 & 493  & 08:25 & M1.4 & 09:12:09 & 280  & 103 \\
    10 & 2022/03/28 & 11:21 & N20W62$^{*}$ &541 & No  & 11:23 & 394  & 10:58 & M4.0 & 12:00:05 & 702  & Halo \\
11 & 2022/03/30 & 17:30 & N18W75$^{*}$ &768 & No  & 17:31 & 769  & 17:21 & X1.3 & 18:00:05 & 641  & Halo \\
12 & 2022/03/31 & 18:27 & N14W61 &790 & No  & 18:34 & 928 & 18:17 & M9.6 & 19:12:05 & 489  & Halo \\
13 & 2022/04/02 & 13:15 & N16W66 &527 & 228 & 13:23 & 680  & 12:56 & M3.9 & 13:36:05 & 1433 & Halo \\
14 & 2022/04/17 & 03:25 & N12E84 &512 & 138 & 03:29 & 520  & 03:17 & X1.1 & 03:48:05 & 895  & 209 \\
15 & 2022/04/20 & 03:55 & S33W93 &652 & No  & 03:55 & 737 & 03:41 & X2.2 & 04:12:05 & 1001 & 139 \\
16 & 2022/05/25 & 18:10 & S20W70$^{*}$ &553 & No  & 18:18 & 592  & 18:09 & M1.3 & 18:36:05 & 1134 & Halo \\
17 & 2022/07/09 & 13:35 & S22W88 &727 & 187 & 13:51 & 784  & 13:29 & C8.5 & 13:48:05 & 1034 & 148 \\
18 & 2022/08/19 & 04:20 & S27W60  &508 & No  & 04:35 & 484  & 04:14 & M1.6 & 04:49:30 & 832  & Halo \\
19 & 2022/08/19 & 20:25 & S26W64 &509 & 66  & 20:32 & 514  & 20:19 & C6.7 & 21:12:09 & 332  & 35  \\
20 & 2022/10/02 & 20:20 & N17W68$^{*}$ &776 & No  & 20:24 & 978 & 19:53 & X1.0 & 20:36:05 & 1086 & Halo \\
21 & 2022/11/09 & 19:58 & S27E62 &514 & No  & 20:03 & 459  & 19:46 & C4.3 & 20:24:05 & 228  & 130 \\
22 & 2022/11/19 & 12:48 & N10W61 &512 & 113 & 12:49 & 458  & 12:42 & M1.6 & 13:25:48 & 422  & 165 \\
23 & 2023/03/03 & 17:53 & N21W76 &530 & 190 & 18:03 & 493  & 17:41 & X2.1 & 18:12:06 & 709  & Halo \\
24 & 2023/03/30 & 07:33 & S22W77 &514 & No  & 07:36 & 473  & 07:24 & M5.4 & 07:48:05 & 487  & 224 \\
25 & 2023/05/05 & 06:58 & N10E62 &675 & 110 & 07:06 & 625  & 06:52 & C9.5 & 08:00:05 & 770  & Halo \\
26 & 2023/07/19 & 17:10 & S18W88 &544 & 207 & 17:12 & 641  & 17:04 & M3.8 & 17:36:06 & 716  & 179 \\
27 & 2023/07/28 & 15:43 & N26W87 &860 & No  & 15:52 & 735  & 15:39 & M4.1 & 16:00:05 & 1896 & Halo \\
28 & 2023/08/05 & 21:52 & N11W77 &662 & 156 & 22:14 & 761  & 21:45 & X1.6 & 22:12:05 & 1647 & Halo \\
29 & 2023/08/07 & 20:32 & N10W102 &533 & No  & 20:38 & 477  & 20:30 & X1.5 & 20:48:05 & 1851 & Halo \\
30 & 2023/08/17 & 02:25 & S23W86 &511 & No & 02:34 & 555  & 02:19 & C3.6 & 02:48:05 & 623  & 239 \\
31 & 2023/09/01 & 02:55 &  N13W73 &510 & No  & 03:08 & 704  & 02:30 & M1.2 & 03:24:05 & 1339 & Halo \\
32 & 2023/09/22 & 19:29 & S21W63 &729 & No  & 19:30 & 657  & 19:20 & C9.8 & 20:12:06 & 221  & 57  \\
33 & 2023/10/26 & 22:55 & N25E90 &519 & No  & 23:23 & 604  & 22:47 & M1.4 & 23:14:06 & 1238 & Halo \\
34 & 2024/01/29 & 04:07 & S28W86 &515 & 172 & 04:08 & 358  & 03:54 & M6.8 & 04:24:05 & 1554 & Halo \\
35 & 2024/02/07 & 03:14 & S34W70 &651 & No  & 03:26 & 722  & 03:04 & M5.1 & 03:24:05 & 1452 & Halo \\
36 & 2024/02/09 & 13:03 & S35W105 &1007& 296 & 13:12 & 968 & 12:53 & X3.3 & 13:25:57 & 2782 & Halo \\
37 & 2024/02/14 & 03:53 & Backside &610 & 145 & 03:55 & 514  & 03:45 & Backside & 04:12:05 & 2191 & Halo \\
38 & 2024/02/16 & 06:53 & S19W82 &758 & No  & 06:55 & 711  & 06:42 & X2.5 & 07:12:05 & 617  & 180 \\
39 & 2024/02/21 & 01:51 & Backside &685 & No  & 02:01 & 545  & 01:40 & Backside & 02:12:05 & 1014 & 188 \\
40 & 2024/05/14 & 17:26 & N17E72 &524 & No  & 17:30 & 417  & 17:25 & M4.4 & 17:48:05 & 1407 & Halo \\
41 & 2024/05/27 & 06:57 & S18E89 &569 & 179 & 07:01 & 483  & 06:49 & X2.8 & 07:24:05 & 1291 & Halo \\
42 & 2024/06/08 & 01:29 & S18W69 &513 & No  & 01:30 & 411  & 01:23 & M9.7 & 01:48:05 & 1427 & Halo \\
43 & 2024/06/10 & 18:20 & S17W106 &618 & No  & 18:22 & 636  & 18:11 & M9.5 & 18:36:05 & 1266 & 249 \\
44 & 2024/06/11 & 22:30 & Backside &706 & No  & 22:35 & 576  & 22:23 & Backside & 23:06:13 & 1988 & Halo \\
45 & 2024/07/03 & 07:37 &S18W61 &908 & No  & 07:44 & 873 & 07:29 & M1.5 & 08:00:05 & 467  & 147 \\
46 & 2024/09/09 & 05:00 & Backside &678 & 259 & 05:10 & 608  & 04:50 & Backside & 05:24:05 & 1522 & Halo \\
47 & 2024/10/24 & 03:36 &  S16E76 &777 & No  & 03:45 & 682  & 03:30 & X3.3 & 03:48:05 & 2385 & Halo \\
48 & 2024/12/05 & 10:38 & S21W90 &607 & 299 & 10:41 & 602  & 10:32 & M2.5 & 11:00:05 & 381  & 111 \\
49 & 2024/12/21 & 19:36 & Backside &662 & 148 & 19:43 & 592  & 19:25 & Backside & 19:48:05 & 607  & 200 \\
50 & 2024/12/23 & 11:12 & S15E61$^{*}$ &641 & 271 & 11:12 & 526  & 11:06 & M8.9 & 11:24:05 & 688  & Halo \\
51 & 2025/04/18 & 23:32 & S26E100 &533 & No  & 23:47 & 532  & 23:08 & M4.5 & 23:48:05 & 1050 & 173 \\
52 & 2025/05/13 & 15:33 & N07W83 &755 & No  & 15:38 & 579  & 15:25 & X1.2 & 15:48:05 & 762  & 139 \\
53 & 2025/08/21 & 07:58 & Backside &709 & No  & 08:22 & 723  & 07:55 & Backside & 08:24:06 & 1612 & Halo \\
54 & 2025/10/30 & 03:58 & Backside &656 & No  & 04:29 & 775  & 03:50 & Backside & 04:24:05 & 1436 & Halo \\
55 & 2025/12/01 & 02:39 & N20E81 &857 & No  & 02:42 & 723 & 02:27 & X1.9 & 02:48:05 & 1810  & Halo \\

\hline
\end{tabular}
\label{table}
\end{table}

\end{document}